\documentclass[prb,twocolumn,secnumarabic,tightenlines,amssymb,nobibnotes,aps]{revtex4}
\usepackage{amssymb}
\usepackage[fleqn]{amsmath}
\usepackage[dvips]{graphicx}
\usepackage{docs}
\usepackage{bm}
\usepackage{textcomp}
\usepackage[colorlinks=true,linkcolor=blue,citecolor=blue,urlcolor=black]{hyperref}%
\expandafter\ifx\csname package@font\endcsname\relax\else
 \expandafter\expandafter
 \expandafter\usepackage
 \expandafter\expandafter
 \expandafter{\csname package@font\endcsname}%
\fi
\newlength{\Figwidth}
\begin{document}
\title{Spin-Splitting and Rashba-Effect at Mono-Layer GaTe in the Presence of Strain}

\author{Mohammad Ariapour{\color{blue}$^{1}$}}
\author{Shoeib~Babaee Touski{\color{blue}$^{2}$}}
\email{touski@hut.ac.ir}
\affiliation{$^{1}$Department of Sciences, Hamedan University of Technology, Hamedan, Iran}
\affiliation{$^{2}$Department of Electrical Engineering, Hamedan University of Technology, Hamedan, Iran}

\begin{abstract}
In this paper, spintronic properties of a mono-layer GaTe under biaxial and uniaxial strain is investigated. Here, spin properties of two structures of GaTe, one with mirror symmetry and the other with inversion 
symmetry, is studied. We have also calculated the band structure of GaTe with and without spin-orbit 
coupling to find out the importance of spin–orbit interaction (SOI) on its band structure. We find 
band gap can be modified by applying spin-orbit coupling in the presence of strain. We explore Mexican-hat dispersion for different structures and different strain. We find Mexican-hat can be tuned by strain however some cases shows any Mexican-hat. We calculate spin-splitting in conduction and valence band in the presence of strain where the structure 
with inversion symmetry doesn't show any splitting. We find in some cases, GaTe indicates Rashba 
dispersion that can be adjusted by strain. The amount of Rashba parameters may be in the order of other reported two-dimensional materials.

\end{abstract}



\maketitle

 \section{introduction}
 Two-dimensional materials have attracted a great attention in the past decade due to their fascinating physical and chemical properties. These excellent properties can result in device applications.  2D layered materials such as Graphene and transition metal di-chalcogenides and others two-dimensional material, can be separated simply by mechanical or liquid-phase exfoliation from their bulk layered solids, which a mono-layer can be exfoliated by the weak interlayer van der Waals (vdW) forces. These mono-layer materials studied extensively for wide range of applications in Field-effect transistors, spintronic devices.  \cite{novoselov2004electric,wang2014black,huang2016plane,guo2015first,fei2014strain}. Nevertheless,  the lack of an intrinsic band gap or high band gap has encountered with problem for application in electronic and optoelectronic devices \cite{li2012graphane,elias2009control,ma2012electronic,zhou2009tuning}.

In which case, the search for new 2D materials that may introduce new properties for specific applications is of technological requirements. Bulk MX(M=Ga,In and X=S, Se,Te) consist of several mono-layers per unit cell weakly bonded by van der Waals forces and form different crystal phases (e.g., $\alpha$ and $\beta$-type) depending on the layer stacking sequence. 

These class of two-dimensional materials has been extensively studied for potential applications 
in photo-detectors, gas sensors and optoelectronic devices \cite{koppens2014photodetectors,hu2014highly,rao2015comparative}. In these material class, GaTe has a special place among the members of the III-VI family, since Ga and Te are massive atoms in which spin
orbit coupling is stronger than the others. In order to become suitable for different applications, 
the corrections of its intrinsic properties is required. For example, a broad range of chemical methods 
\cite{csahin2011structural} and doping methods\cite{sheng2012synthesis} have been studied to tune the 
band gap of GaTe. Gallium telluride (GaTe), can be an ideal candidate for using strain engineering 
as an effective method to change their electronic and spin properties. It has been proved that 
the band gap of GaTe can be controlled by applying strain both experimentally\cite{cocco2010gap} and theoretically\cite{gui2008band}.

Mexican hat dispersion in this class have been studied and It has been studied that a Mexican-hat-shaped dispersion observed at the valence band maximum (VBM) in monolayer of GaTe. This dispersion in the band structure lead to van Hove singularity in the density of states (DOS). Mexican hat dispersion provides 1D-like electronic density of states of the valence band near the band edge \cite{rybkovskiy2014transition}. The effects of this singularity on the magnetic and electronic properties has been studied \cite{wickramaratne2014electronic,cao2015tunable,zolyomi_band_2013,wu2014magnetisms,magorrian2016electronic,
rybkovskiy2014transition}. However, this sharp van Hove singularity near the Fermi level could induce a spin splitting in the band structure and suitable magnetism\cite{cao2015tunable,wu2014magnetisms}. One can observe Lifshitz phase transition due to this sharp van Hove singularity \cite{lifshitz1960anomalies}.The high density of states at the valence band edge, in principle, is beneficial to high thermoelectric power factor in this material \cite{wickramaratne2015electronic}.

Spin-orbit-induced spin splitting and spin-orbit interaction (SOI) is believed be responsible of Spin relaxation mechanisms in 2D materials. Dat T. Do et.al. \cite{do2015spin} studied  spin-splitting in 2D mono chalcogenides. They considered GaTe in monoclinic structure and reported zero spin splitting because it has inversion symmetry  at mono-layer.

Here, we study two different phases of GaTe and observe spin splitting in one phase and other phase exhibits any spin-splitting due to the inversion symmetry. Spin-splitting can be tuned by tensile and compressive strain. Rashba effect also observed in compressive regime. Rashba spin-orbit interaction originates from the lack of inversion symmetry \cite{bychkov1984properties}. 

The Rashba effect has been studied intensively in the spintronics since it can be used to tune the spin direction by means of an electric field and applicable in spin-field effect transistor \cite{datta1990electronic}. Experimentally, Rashba effect has been reported experimentally in quantum wells , surfaces of heavy metals, and BiTeI and theoretical for two-dimensional materials, for instance graphene, polar transition-metal dichalcogenide monolayers \cite{absor2018tunable}, the MoS2 /Bi(111)
heterostructure, GaX(mono-chalcogenides)/TMDC heterostructures \cite{zhang2018rashba}, and gated multilayer InSe \cite{premasiri2018tuning}. Spin properties is the focus of the present study \cite{nitta1997gate, lashell1996s,
vzutic2004vzutic, kimura2010strong, konschuh2010tight, cheng2013spin, liu2013tunable, min2006intrinsic,
ishizaka2011k, ast2007giant, koroteev2004strong, lee2015giant}.

In spintronics and optoelectronics have stimulated the research for materials capable of presenting a 
great degree of spin polarization and long spin relaxation time \cite{gui2008band11}. If symmetry is 
broken by strain, quantum phase transitions such as semiconductor-semimetal transition are possible
\cite{zhang2015atomically}.

In this work we discuss the magnetic and electronic properties of monolayer crystals of GaTe. We present the electronic band structures and spin-splitting for monolayer Gallium telluride applying by biaxial and uniaxial in-plane strain.

\section{Computational details}

 The analysis of monolayer GaTe was carried out using density-functional theory (DFT) as implemented in the Quantum Espresso \cite{giannozzi2009quantum,giannozzi2017advanced} plane-wave-basis codes using projector augmented wave
(PAW) \cite{blochl94,kresse99}. To calculate the geometries and band-structures, we used semilocal exchange-correlation functional functional with high-throughput ultrasoft
pseudopotential: PBEsol \cite{perdew2008restoring}. The plane-wave cutoff energy was 612 eV. During calculation, a $10\times 10 \times 1$ Monkhorst-Pack k-point grid was used for electrical properties but a denser k-mesh should be used for Rashba calculations, here we used a $16\times16 \times 1$ k-point mesh. We were first performed full geometry optimization until the forces on the atoms are less than 0.01 $\mathrm{eV/\AA}$. 

We explore two phases  $\alpha$ and $\beta$ of GaTe monolayer which these two phases was shown in the Fig. \ref{Fig:abband}. $\alpha$ phase with $D_{3h}$ group symmetry forms a 2D honeycomb structure, which vertically placed Ga and Te pairs at two different sub-lattices. The structure of $\beta$ with $D_{3d}$ phase can be observed in  of Fig. \ref{Fig:abband}(b), that one of the Te layer shifted with respect to the other. This phase breaks the mirror symmetry of the original structure but forms inversion symmetry. 

Biaxial and uniaxial strain is applied for each phases and geometry is completely relaxed for two phases. Total energy and lattice constant for two phases is the same. Band structure with and without spin-orbit coupling is studied electronic and spintronic properties is investigated.
 
\section{Results and discussion}

\begin{figure}[!ht]
\begin{center}
\includegraphics[width=1.0\linewidth]{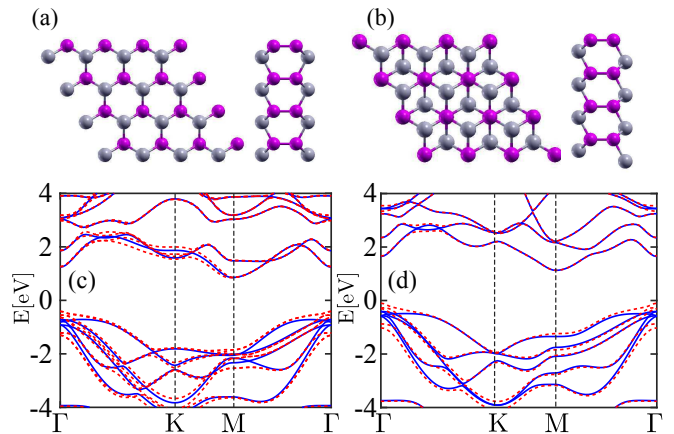}
\end{center}
\caption{Schematic of mono-layer GaTe for $\alpha$ (a) and $\beta$ phases (b) structures. Band structure of gallium telluride in phase $\alpha$ (a) and $\beta$ (b) without(blue lines) 
and with (red dash-lines) spin-orbit coupling.} 
\label{Fig:abband}
\end{figure}

The calculated electronic band structures for two phases are plotted in the Fig. \ref{Fig:abband}. Two phases are indirect-gap semiconductors, primarily due to the valence-band maximum (VBM) move from the $\Gamma$ point and the conduction-band minimum (CBM) located at the $M$ point. The band-gap in the phase $\alpha$ is $E_{g}=1.57eV$. The indirect band gap in phase $\alpha$ is in agreement with the DFT results 
\cite{ge2017first,cai2016exciton,cai2016band,huang2015effects,zhuang2013single} for monolayer GaTe.
Although, we calculated the electronic properties of GaTe with the spin-orbit coupling $(SOC)$ which Fig. \ref{Fig:abband} indicates SOC changes the band structure and then in this regime, band-gap declines to $E_{g}=1.256eV$. SOC consideration causes to band-gap deceases by about $0.3 eV$. Spin-orbit interaction also moves the valence-band maximum to $\Gamma$ point and removes Mexican-hat dispersion.

$\beta$ phase of GaTe is the other structure, that can be distinguished by comparing optically active 
[infrared (IR) and Raman] phonon spectra. We find that the band-gap in this phase, similar to $\alpha$ phase, is indirect and is $E_{g}=1.39eV$ which  band-gap decreases to $E_{g}=1.228eV$ by applying SOC. Band-gap decreases approximately $0.16 eV$ which is lower than $\alpha$ phase. Therefore, the effects of SOC consideration in $\beta$ phase  is weaker than $\alpha$ phase. This can be implied two spin components are degenerated due to inversion symmetry.

\begin{figure}[!ht]
\begin{center}
\includegraphics[width=1.0\linewidth]{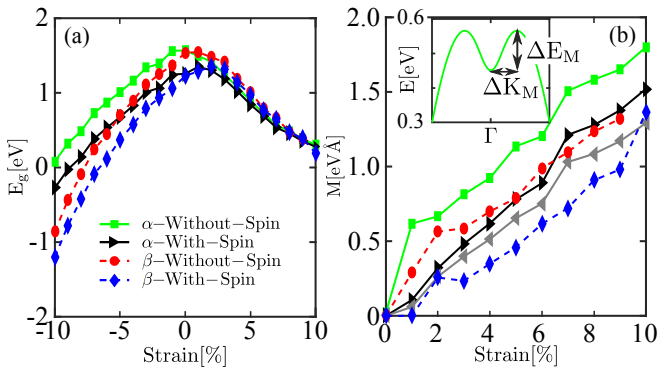}
\end{center}
\caption{(a) The band gap and (b) the Mexican-hat of GaTe in phase $\alpha$ and phase $\beta$ with applying biaxial strain respect the strain.} 
\label{Fig:abband-strain}
\end{figure}

The band-gap value of GaTe can be further tuned by the external factors, such as the bias voltage or strains. The strain modulates band-gap of GaTe monolayer as are shown in the Fig. \ref{Fig:abband-strain}. The homogeneous in-plane strains are applied to the monolayer by varying the lattices constant as $(l-l_{0})/l_{0}$, where $l(l_{0})$  is the lattice constant under the strain (equilibrium) condition.

The variation range of strain is $-10\%$-$10\%$ from equilibrium condition. In the compressive strain regime, we find that although the horizontal distance of Ga-Te is shortened, the buckling heights in the vertical direction are increased for the GaTe monolayer. The charge redistribution is weaken duration the compressive strain and the band-gap is decreased. In this strain range, although CBM is in the M point, a strain-induced self-doping occurs due to the movement of bands at the $\Gamma$ point. In the tensile strain regime, though the horizontal distance of Ga-Te is increased and  the buckling heights in the vertical direction is increased for the GaTe monolayer. The charge redistribution is strength duration the tensile strain and the band-gap decreases. Band-gap for two phases approximately is the same in tensile regime and differences between these two band-gaps increases by moving from tensile to compressive regime. $\alpha$ phase relative to $\beta$ phase has a higher band-gap in compressive regime. Therefore, the effect of strain in $\alpha$ phase is lower than $\beta$ phase.

SOC in $\alpha$ phase is similar to $\beta$ phase, for example, in the $-10 \%$ strain, the differences between band-gap with and without SOC in $\alpha$ and $\beta$ phases is $0.353$ and $0.3417$, respectively. Another result of Fig. \ref{Fig:abband-strain} is due to the strain, GaTe in phase $\beta$ could be phase transition from semiconductor to semi-metal in strain $-10$ whereas $\alpha$ phase remains semiconductor in this strain.

The valence band maximum shifted to $\Gamma$ point by the compressive strain and dismiss Mexican-hat dispersion. At the other hand, the tensile strain regime shows the Mexican-hat-shaped dispersion \cite{cao2015tunable} in the band structure and increases the density of state in valance bands (Fig. \ref{Fig:abband-strain}). We obtained Mexican-hat coefficient with SOC for both phases in 
up and down components spin whereas in $\beta$ phase these two components are degenerated. Mexican-hat coefficient can be defined as: $M=\Delta E_M/k_M$ where $\Delta E_M$ and $k_M$ are energy and momentum differences between VBM and $\Gamma$-point, respectively, see the inset of Fig. \ref{Fig:abband-strain}(b). In $\alpha$ phase, Mexican-hat has higher value than the $\beta$ phase. SOI decreases Mexican-hat 
coefficient for two phases. For example, in $5 \%$ strain, SOC decreases $42 \%$ Mexican-hat coefficient which this decreasing ratio approximately hold for other percentage of strain in tensile regime. As can be seen from the Fig. \ref{Fig:abband-strain}, up-spin component of $\alpha$ phase shows higher Mexican-hat relative to down component and their difference increases with increment of strain.

\begin{figure}[!ht]
\begin{center}
\includegraphics[width=1.0\linewidth]{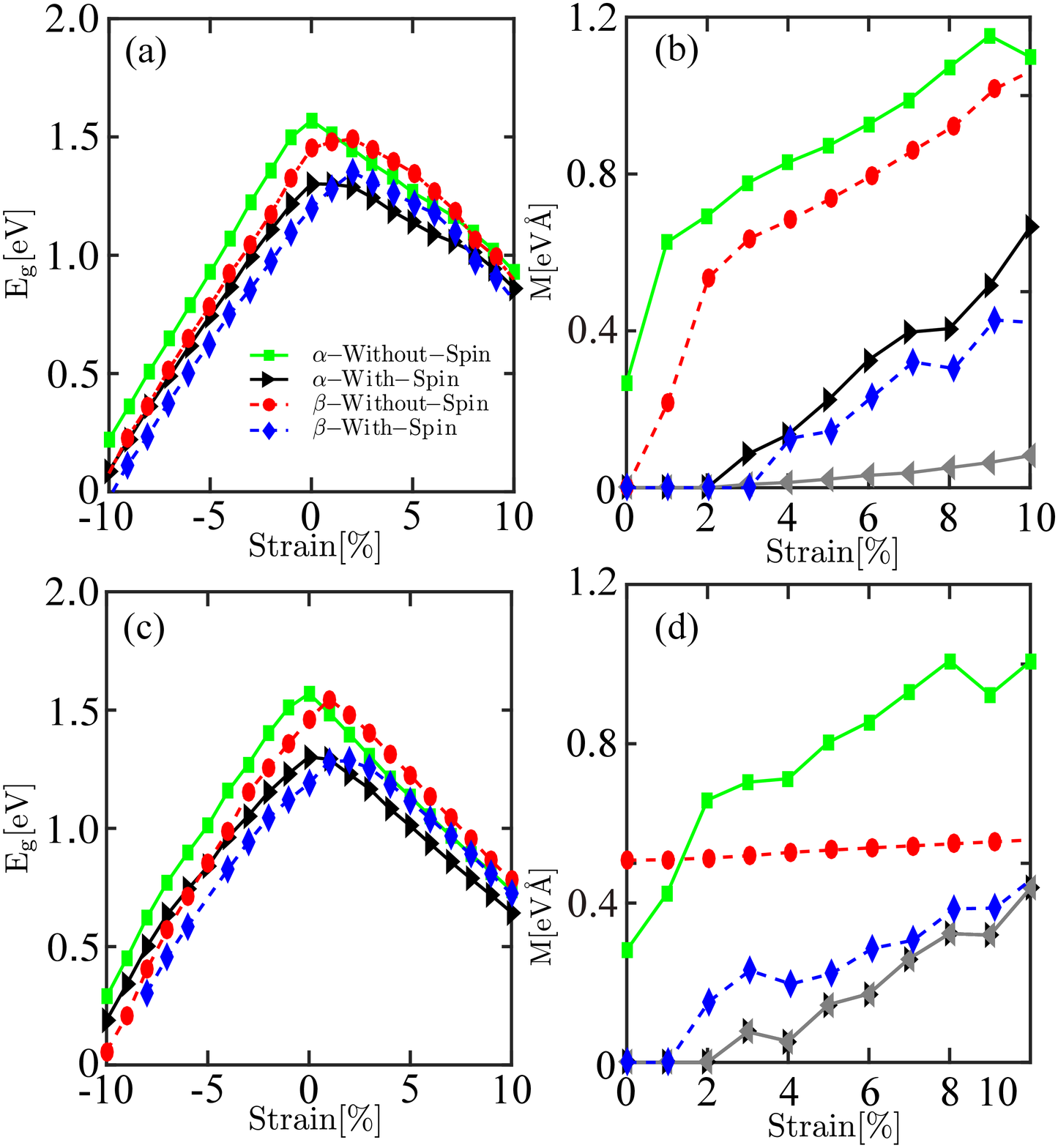}
\end{center}
\caption{The band gap and the Mexican-hat of GaTe in phase $\alpha$ and $\beta$ with applying Armchair (a,b)and Zigzag 
(c,d) uniaxial strains.} 
\label{Fig:eg-mex-uni}
\end{figure}

We study the band gap and Mexican-hat dispersion with applying uniaxial strain in two Armchair and Zigzag directions that is shown in the Fig. \ref{Fig:eg-mex-uni}. Our finding indicates that the gap with- and without-SOC changes respect to the strain and shows the same behavior in both directions. Band gap decreases with increasing strain in compression regime. In the tensile strain regime,  band gap decreases by increasing strain. For both directions and both structures ($\alpha$ and $\beta$), SOC declines band gap relative to without SOC.

The Mexican-hat coefficient increases with increasing strain in both Armchair and Zigzag strains. 
In the $\alpha$ structure, Mexican-hat is higher than the $\beta$ structure. SOC declines Mexican-hat for both strain directions in two structures. For example, in $\beta$ structure, SOC approximately decreases Mexican-hat by 4 times, Whereas, SOC decreases this coefficient lower than 2 times for biaxial strain. Mexican-hat coefficient is calculated for two spins (up and down). For $\beta$ structure these two coefficients is the same because of spin degeneracy in this structure due to inversion symmetry. $\alpha$ structure with no inversion symmetry, shows two Mexican-hats for both  separated spins. By applying Armchair strain, VBM is located at $\Gamma-M$ direction and there isn't any spin-splitting due to spin degeneracy in this direction. At the other hand, MVB is placed in $\Gamma-K$ direction and one can obtain two Mexican-hat coefficients for both spins.

\begin{figure}[!ht]
\begin{center}
\includegraphics[width=1.0\linewidth]{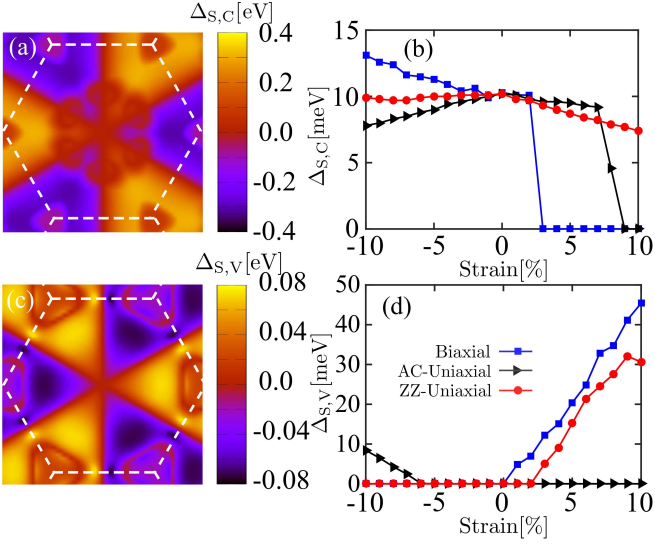}
\end{center}
\caption{Spin Splitting of (a)the conduction and (c)the valance bands in the first Brillouin zone. Spin-splitting at (b) CBM and (d) VBM as a function of strain in phase $\alpha$ with applying Biaxial, Armchair and 
Zigzag uniaxial strains.} 
\label{Fig:spsp}
\end{figure}

Fig. \ref{Fig:spsp} shows the calculations of the spin-splitting of the uppermost valence band $(\Delta_{S,V})$ and the lowermost conduction band $(\Delta_{S,C})$ in monolayer GaTe for phase $\alpha$ at biaxial, Armchair and Zigzag uniaxial strains. GaTe in phase $\beta$ doesn't induce spin-splitting due to the inversion symmetry. Spin-splitting in conduction and valence bands at Brillouin zone is reported in the Fig \ref{Fig:spsp} (a) and (c), respectively. In the conduction band, between M point to M' point (area 1) spin-splitting is positive whereas between M' to M point (area 2) is negative.  This indicates the up-spin energy is higher than the down-spin energy at area 1, and vice versa at area 2. Spin-splitting, for the valence band,  at area 1 is negative and at area 2 is positive. $\Delta_{S,C}$ decreases by increasing biaxial strain. For strain higher than $2 \%$, CBM is located at $\Gamma$ point and there isn't any spin-splitting. However, for lower than $2 \%$, CBM is placed at  $M$ point. In compressive regime, as Ga-Te atom distances decreases, increasing of potential gradient causes to SOC increasing $(\lambda_{s} \propto \nabla V)$ \cite{}. VBM is located at $\Gamma$ point for compressive strain then $\Delta_{S,V}$  vanished. Whereas, VBM is located at $\Gamma$ point for all strains in Armchair direction, therefore spin-splitting goes to zero. Due to VBM move to $\Gamma^{*}$ and take a distance from $\Gamma$ point. $k_{0}$ vector can be defined as distance between $\Gamma$ and $\Gamma^{*}$. $k_{0}$ increases by tensile strain increasing and $\Delta_{S,V}$ increases as $k_{0}$ increasing.

\begin{figure}[!ht]
\begin{center}
\includegraphics[width=0.7\linewidth]{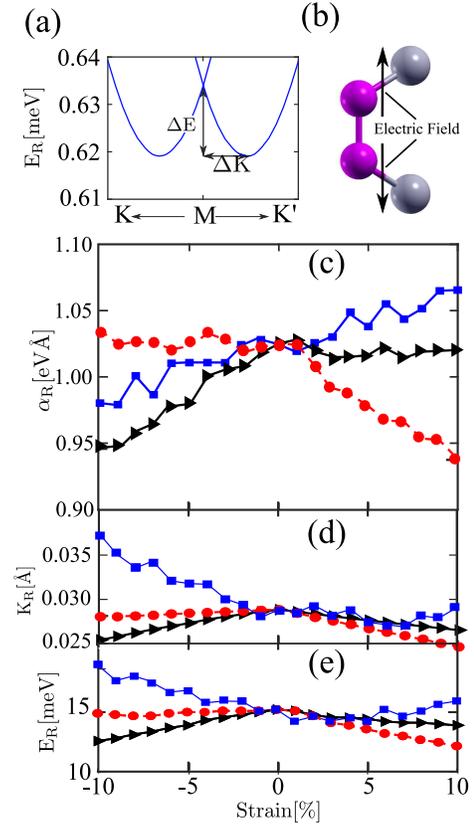}
\end{center}
\caption{(a) Band structure in the direction that The Rashba effect can happen. (b) The unit cell of GaTe with effective electric field. Rashba parameter (c), Rashba energy (d) and Rashba momentum of the conduction band in the phase $\alpha$ with applying Biaxial, Armchair and Zigzag uniaxial 
strains.} 
\label{Fig:rashba}
\end{figure}

Rashba effect can play an important role in future spintronic. The Rashba effect happens due to the structural asymmetry causes an out-of-plane electric field whereas, GaTe with mirror symmetry shouldn't show any Rashba effect but it constructed from two sub-layers that each sub-layer has an out-of-plane asymmetry. An electric field is originated from this asymmetry, see Fig. \ref{Fig:rashba}. For a two-dimensional electron gas the Rashba Hamiltonian is defined by:
\begin{equation}
 H_{R} = \alpha_{R} \hat {\sigma}. (\hat{k} \times \hat{E_{z}})
 \label{eqRashba}
\end{equation}
Where $\alpha_{R}$ is the Rashba parameter, $\hat{\sigma}$ is the Pauli matrices, $\hat{k}$ is the in-plane momentum of the electrons, and $\hat{E}_{z}$ is the out-of-plane unit vector \cite{bychkov1984properties}. $\alpha_{R}$ can be defined as $\alpha_{R} = 2E_{R}/K_{R}$ where $E_{R}$ is Rashba energy and $k_{R}$ is momentum offset, see Fig.\ref{Fig:rashba}(a). However, $E_{R}$ and $k_{R}$  stabilize spin precession and make a phase offset for different spin channels in the spin-field effect transistor device \cite{absor2018strong}. Rashba-split dispersion for $\alpha$ phase is reported in Fig. \ref{Fig:rashba}. This figure shows $\alpha_{R}$ increases with increasing the strain in biaxial strain that is in agreement with Ref.[\onlinecite{ma2014emergence}]. By increasing strain, distance between Ga sublayers and Te sublayers decreases which causes to enhance out-of-plane electric field. Rashba parameter increases in the Armchair strain in compressive regime and reaches to a constant amount in tensile regime. In compressive regime Armchair strain declines these two sub-layers. Whereas, Zigzag strain is in the opposite of Armchair strain. 

One can observe the results of the parameters $E_R$, $k_R$, and $\alpha_R$ in Table. \ref{table:rashba}, and compare with some  materials for previously reported references.  We find the amount of $\alpha_{R}$ is 1.03 for equilibrium condition and 1.07 for $10 \%$ strain. However, $\alpha_{R}$ is the largest calculated value under biaxial strain, but in this strain, CBM is located at $\Gamma$ point.  For the comparison with the other papers, Rashba parameters is obtained 1.03 for no strain which in this strain CBM is placed at M point. Our $\alpha_{R}$ is larger than GaSe/MoSe$_2$ heterostructure \cite{zhang2018rashba}, Janus Transition Metal Dichalcogenides \cite{li2017electronic,yao2017manipulation}, comparable with Br-doped PtSe$_2$ \cite{absor2018strong} and lower than LaOBiS$_2$ \cite{liu2013tunable}, BiSb \cite{singh2017giant},BiTeI \cite{ma2014emergence},
I-doped PtSe$_2$ \cite{absor2018strong}. $E_R$ indicates potential barrier between two spin. 
$E_R$ in the strained GaTe is lower than Bi/Ag(111) interface, LaOBiS$_2$ and BiTeI. However, monolayer WSeTe and GaSe/MoS$_2$ heterostructure show higher $E_R$ but strained GaTe shows higher and comparable $E_R$ relative some reported previously. For instance, GaTe shows higher amount relative to I- and B-doped PtSe$_2$, WSSe and MoSSe. These results approves GaTe can be a good candidate for future spintronic.

\begin{table}[t]
\caption{Parameters of the Rashba splitting for different materials: the momentum offset $K_{R}[\AA^{-1}]$, Rashba
energy $E_{R}[meV]$ and Rashba parameter $\alpha_{R}[eV\AA]$.}
\centering 
\begin{tabular}{l c c c} 
\hline\hline 
 & $E_{R}$ & $K_{R}$ & $\alpha_{R}$ \\ [0.5ex]
\hline 
2D-ML & & & \\
 GaTe ($0 \%$ ) & 14.9 & 0.0288 & 1.03\\ 
GaTe ($10 \%$) & 15.5 & 0.029 & 1.07 \\ 
LaOBiS$_2$ \cite{liu2013tunable} & 38 & 0.025 & 3.04 \\
BiSb \cite{singh2017giant} & 13 & 0.0113 & 2.3 \\
BiTeI \cite{ma2014emergence} & 39.8 & 0.043 & 1.86 \\
I-doped PtSe$_2$ \cite{absor2018strong} & 12.5 & 0.015 & 1.7 \\
Br-doped PtSe$_2$ \cite{absor2018strong} & 4.91 & $9.15 \times 10^{-3}$ & 1.07 \\
WSeTe \cite{yao2017manipulation} & 52 & 0.17 & 0.92 \\
WSSe  \cite{li2017electronic} & 3.6 & 0.010 & 0.72 \\
MoSSe \cite{li2017electronic} & 1.4 & 0.005 & 0.53 \\
GaSe/MoSe$_2$ heterostructure \cite{zhang2018rashba} & 31 & 0.13 & 0.49 \\
Surface & & & \\
Bi/Ag (111) \cite{ast2007giant} & 200 & 0.13 & 3.05 \\

Bi (111)  \cite{koroteev2004strong}& 14 & 0.05 & 0.55 \\
Interface & & & \\
InGaAs/InAlAs  \cite{nitta1997gate}& $<1$ & 0.028 & 0.07 \\ 
[1ex] 
\hline 
\end{tabular}
\label{table:rashba} 
\end{table}


In conclusion, using first-principles density functional calculations, we have investigated systematically the spintronic properties in two-dimensional GaTe. Two structures, $\alpha$ with mirror symmetry and $\beta$ with inversion symmetry is considered. This study suggests band gap can be tuned under biaxial, AC-uniaxial and ZZ-uniaxial strains with and without spin-orbit coupling. SOI causes to band gap shows lower amount. In the tensile and compressive strain, the band gaps of GaTe decreased by increasing strain amount.  Although, the band gap more decreases by consideration of SOC. The AC-strain and ZZ-strain apply a similar effect on the band gap of GaTe. We observe Mexican-hat only in the case of tensile strain and Mexican-hat increases with strain increasing. SOI implies to smoother Mexican-hat dispersion. In biaxial strain, The spin-splitting declines with increment of strain. GaTe indicates Rashba effect that Rashba parameters can be tuned a little by strain. 



\begin{thebibliography}{59}
	\expandafter\ifx\csname natexlab\endcsname\relax\def\natexlab#1{#1}\fi
	\expandafter\ifx\csname bibnamefont\endcsname\relax
	\def\bibnamefont#1{#1}\fi
	\expandafter\ifx\csname bibfnamefont\endcsname\relax
	\def\bibfnamefont#1{#1}\fi
	\expandafter\ifx\csname citenamefont\endcsname\relax
	\def\citenamefont#1{#1}\fi
	\expandafter\ifx\csname url\endcsname\relax
	\def\url#1{\texttt{#1}}\fi
	\expandafter\ifx\csname urlprefix\endcsname\relax\def\urlprefix{URL }\fi
	\providecommand{\bibinfo}[2]{#2}
	\providecommand{\eprint}[2][]{\url{#2}}
	
	\bibitem[{\citenamefont{Novoselov et~al.}(2004)\citenamefont{Novoselov, Geim,
			Morozov, Jiang, Zhang, Dubonos, Grigorieva, and
			Firsov}}]{novoselov2004electric}
	\bibinfo{author}{\bibfnamefont{K.~S.} \bibnamefont{Novoselov}},
	\bibinfo{author}{\bibfnamefont{A.~K.} \bibnamefont{Geim}},
	\bibinfo{author}{\bibfnamefont{S.~V.} \bibnamefont{Morozov}},
	\bibinfo{author}{\bibfnamefont{D.}~\bibnamefont{Jiang}},
	\bibinfo{author}{\bibfnamefont{Y.}~\bibnamefont{Zhang}},
	\bibinfo{author}{\bibfnamefont{S.~V.} \bibnamefont{Dubonos}},
	\bibinfo{author}{\bibfnamefont{I.~V.} \bibnamefont{Grigorieva}},
	\bibnamefont{and} \bibinfo{author}{\bibfnamefont{A.~A.}
		\bibnamefont{Firsov}}, \bibinfo{journal}{Science}
	\textbf{\bibinfo{volume}{306}}, \bibinfo{pages}{666} (\bibinfo{year}{2004}).
	
	\bibitem[{\citenamefont{Wang et~al.}(2014)\citenamefont{Wang, Wang, Xia, Wang,
			Jiang, Xia, Chin, Dubey, and Han}}]{wang2014black}
	\bibinfo{author}{\bibfnamefont{H.}~\bibnamefont{Wang}},
	\bibinfo{author}{\bibfnamefont{X.}~\bibnamefont{Wang}},
	\bibinfo{author}{\bibfnamefont{F.}~\bibnamefont{Xia}},
	\bibinfo{author}{\bibfnamefont{L.}~\bibnamefont{Wang}},
	\bibinfo{author}{\bibfnamefont{H.}~\bibnamefont{Jiang}},
	\bibinfo{author}{\bibfnamefont{Q.}~\bibnamefont{Xia}},
	\bibinfo{author}{\bibfnamefont{M.~L.} \bibnamefont{Chin}},
	\bibinfo{author}{\bibfnamefont{M.}~\bibnamefont{Dubey}}, \bibnamefont{and}
	\bibinfo{author}{\bibfnamefont{S.-j.} \bibnamefont{Han}},
	\bibinfo{journal}{Nano Lett.} \textbf{\bibinfo{volume}{14}},
	\bibinfo{pages}{6424} (\bibinfo{year}{2014}).
	
	\bibitem[{\citenamefont{Huang et~al.}(2016)\citenamefont{Huang, Tatsumi, Ling,
			Guo, Wang, Watson, Puretzky, Geohegan, Kong, Li et~al.}}]{huang2016plane}
	\bibinfo{author}{\bibfnamefont{S.}~\bibnamefont{Huang}},
	\bibinfo{author}{\bibfnamefont{Y.}~\bibnamefont{Tatsumi}},
	\bibinfo{author}{\bibfnamefont{X.}~\bibnamefont{Ling}},
	\bibinfo{author}{\bibfnamefont{H.}~\bibnamefont{Guo}},
	\bibinfo{author}{\bibfnamefont{Z.}~\bibnamefont{Wang}},
	\bibinfo{author}{\bibfnamefont{G.}~\bibnamefont{Watson}},
	\bibinfo{author}{\bibfnamefont{A.~A.} \bibnamefont{Puretzky}},
	\bibinfo{author}{\bibfnamefont{D.~B.} \bibnamefont{Geohegan}},
	\bibinfo{author}{\bibfnamefont{J.}~\bibnamefont{Kong}},
	\bibinfo{author}{\bibfnamefont{J.}~\bibnamefont{Li}}, \bibnamefont{et~al.},
	\bibinfo{journal}{ACS Nano} \textbf{\bibinfo{volume}{10}},
	\bibinfo{pages}{8964} (\bibinfo{year}{2016}).
	
	\bibitem[{\citenamefont{Guo et~al.}(2015)\citenamefont{Guo, Wang, Kuang, and
			Huang}}]{guo2015first}
	\bibinfo{author}{\bibfnamefont{R.}~\bibnamefont{Guo}},
	\bibinfo{author}{\bibfnamefont{X.}~\bibnamefont{Wang}},
	\bibinfo{author}{\bibfnamefont{Y.}~\bibnamefont{Kuang}}, \bibnamefont{and}
	\bibinfo{author}{\bibfnamefont{B.}~\bibnamefont{Huang}},
	\bibinfo{journal}{Phys. Rev. B} \textbf{\bibinfo{volume}{92}},
	\bibinfo{pages}{115202} (\bibinfo{year}{2015}).
	
	\bibitem[{\citenamefont{Fei and Yang}(2014)}]{fei2014strain}
	\bibinfo{author}{\bibfnamefont{R.}~\bibnamefont{Fei}} \bibnamefont{and}
	\bibinfo{author}{\bibfnamefont{L.}~\bibnamefont{Yang}},
	\bibinfo{journal}{Nano Lett.} \textbf{\bibinfo{volume}{14}},
	\bibinfo{pages}{2884} (\bibinfo{year}{2014}).
	
	\bibitem[{\citenamefont{Li et~al.}(2012)\citenamefont{Li, Li, and
			Chen}}]{li2012graphane}
	\bibinfo{author}{\bibfnamefont{Y.}~\bibnamefont{Li}},
	\bibinfo{author}{\bibfnamefont{F.}~\bibnamefont{Li}}, \bibnamefont{and}
	\bibinfo{author}{\bibfnamefont{Z.}~\bibnamefont{Chen}}, \bibinfo{journal}{J.
		Am. Chem. Soc.} \textbf{\bibinfo{volume}{134}}, \bibinfo{pages}{11269}
	(\bibinfo{year}{2012}).
	
	\bibitem[{\citenamefont{Elias et~al.}(2009)\citenamefont{Elias, Nair,
			Mohiuddin, Morozov, Blake, Halsall, Ferrari, Boukhvalov, Katsnelson, Geim
			et~al.}}]{elias2009control}
	\bibinfo{author}{\bibfnamefont{D.~C.} \bibnamefont{Elias}},
	\bibinfo{author}{\bibfnamefont{R.~R.} \bibnamefont{Nair}},
	\bibinfo{author}{\bibfnamefont{T.}~\bibnamefont{Mohiuddin}},
	\bibinfo{author}{\bibfnamefont{S.}~\bibnamefont{Morozov}},
	\bibinfo{author}{\bibfnamefont{P.}~\bibnamefont{Blake}},
	\bibinfo{author}{\bibfnamefont{M.}~\bibnamefont{Halsall}},
	\bibinfo{author}{\bibfnamefont{A.}~\bibnamefont{Ferrari}},
	\bibinfo{author}{\bibfnamefont{D.}~\bibnamefont{Boukhvalov}},
	\bibinfo{author}{\bibfnamefont{M.}~\bibnamefont{Katsnelson}},
	\bibinfo{author}{\bibfnamefont{A.}~\bibnamefont{Geim}}, \bibnamefont{et~al.},
	\bibinfo{journal}{Science} \textbf{\bibinfo{volume}{323}},
	\bibinfo{pages}{610} (\bibinfo{year}{2009}).
	
	\bibitem[{\citenamefont{Ma et~al.}(2012)\citenamefont{Ma, Dai, Guo, Niu, Zhang,
			and Huang}}]{ma2012electronic}
	\bibinfo{author}{\bibfnamefont{Y.}~\bibnamefont{Ma}},
	\bibinfo{author}{\bibfnamefont{Y.}~\bibnamefont{Dai}},
	\bibinfo{author}{\bibfnamefont{M.}~\bibnamefont{Guo}},
	\bibinfo{author}{\bibfnamefont{C.}~\bibnamefont{Niu}},
	\bibinfo{author}{\bibfnamefont{Z.}~\bibnamefont{Zhang}}, \bibnamefont{and}
	\bibinfo{author}{\bibfnamefont{B.}~\bibnamefont{Huang}},
	\bibinfo{journal}{Phys. Chem. Chem. Phys.} \textbf{\bibinfo{volume}{14}},
	\bibinfo{pages}{3651} (\bibinfo{year}{2012}).
	
	\bibitem[{\citenamefont{Zhou et~al.}(2009)\citenamefont{Zhou, Wu, Zhou, and
			Sun}}]{zhou2009tuning}
	\bibinfo{author}{\bibfnamefont{J.}~\bibnamefont{Zhou}},
	\bibinfo{author}{\bibfnamefont{M.~M.} \bibnamefont{Wu}},
	\bibinfo{author}{\bibfnamefont{X.}~\bibnamefont{Zhou}}, \bibnamefont{and}
	\bibinfo{author}{\bibfnamefont{Q.}~\bibnamefont{Sun}},
	\bibinfo{journal}{Appl. Phys. Lett.} \textbf{\bibinfo{volume}{95}},
	\bibinfo{pages}{103108} (\bibinfo{year}{2009}).
	
	\bibitem[{\citenamefont{Koppens et~al.}(2014)\citenamefont{Koppens, Mueller,
			Avouris, Ferrari, Vitiello, and Polini}}]{koppens2014photodetectors}
	\bibinfo{author}{\bibfnamefont{F.}~\bibnamefont{Koppens}},
	\bibinfo{author}{\bibfnamefont{T.}~\bibnamefont{Mueller}},
	\bibinfo{author}{\bibfnamefont{P.}~\bibnamefont{Avouris}},
	\bibinfo{author}{\bibfnamefont{A.}~\bibnamefont{Ferrari}},
	\bibinfo{author}{\bibfnamefont{M.}~\bibnamefont{Vitiello}}, \bibnamefont{and}
	\bibinfo{author}{\bibfnamefont{M.}~\bibnamefont{Polini}},
	\bibinfo{journal}{Nat. Nanotechnol.} \textbf{\bibinfo{volume}{9}},
	\bibinfo{pages}{780} (\bibinfo{year}{2014}).
	
	\bibitem[{\citenamefont{Hu et~al.}(2014)\citenamefont{Hu, Zhang, Yoon, Qiao,
			Zhang, Feng, Tan, Zheng, Liu, Wang et~al.}}]{hu2014highly}
	\bibinfo{author}{\bibfnamefont{P.}~\bibnamefont{Hu}},
	\bibinfo{author}{\bibfnamefont{J.}~\bibnamefont{Zhang}},
	\bibinfo{author}{\bibfnamefont{M.}~\bibnamefont{Yoon}},
	\bibinfo{author}{\bibfnamefont{X.-F.} \bibnamefont{Qiao}},
	\bibinfo{author}{\bibfnamefont{X.}~\bibnamefont{Zhang}},
	\bibinfo{author}{\bibfnamefont{W.}~\bibnamefont{Feng}},
	\bibinfo{author}{\bibfnamefont{P.}~\bibnamefont{Tan}},
	\bibinfo{author}{\bibfnamefont{W.}~\bibnamefont{Zheng}},
	\bibinfo{author}{\bibfnamefont{J.}~\bibnamefont{Liu}},
	\bibinfo{author}{\bibfnamefont{X.}~\bibnamefont{Wang}}, \bibnamefont{et~al.},
	\bibinfo{journal}{Nano Research} \textbf{\bibinfo{volume}{7}},
	\bibinfo{pages}{694} (\bibinfo{year}{2014}).
	
	\bibitem[{\citenamefont{Rao et~al.}(2015)\citenamefont{Rao, Gopalakrishnan, and
			Maitra}}]{rao2015comparative}
	\bibinfo{author}{\bibfnamefont{C.}~\bibnamefont{Rao}},
	\bibinfo{author}{\bibfnamefont{K.}~\bibnamefont{Gopalakrishnan}},
	\bibnamefont{and} \bibinfo{author}{\bibfnamefont{U.}~\bibnamefont{Maitra}},
	\bibinfo{journal}{ACS Appl. Mater. Interfaces} \textbf{\bibinfo{volume}{7}},
	\bibinfo{pages}{7809} (\bibinfo{year}{2015}).
	
	\bibitem[{\citenamefont{{\c{S}}ahin and Ciraci}(2011)}]{csahin2011structural}
	\bibinfo{author}{\bibfnamefont{H.}~\bibnamefont{{\c{S}}ahin}} \bibnamefont{and}
	\bibinfo{author}{\bibfnamefont{S.}~\bibnamefont{Ciraci}},
	\bibinfo{journal}{Phys. Rev. B} \textbf{\bibinfo{volume}{84}},
	\bibinfo{pages}{035452} (\bibinfo{year}{2011}).
	
	\bibitem[{\citenamefont{Sheng et~al.}(2012)\citenamefont{Sheng, Gao, Bao, Wang,
			and Xia}}]{sheng2012synthesis}
	\bibinfo{author}{\bibfnamefont{Z.-H.} \bibnamefont{Sheng}},
	\bibinfo{author}{\bibfnamefont{H.-L.} \bibnamefont{Gao}},
	\bibinfo{author}{\bibfnamefont{W.-J.} \bibnamefont{Bao}},
	\bibinfo{author}{\bibfnamefont{F.-B.} \bibnamefont{Wang}}, \bibnamefont{and}
	\bibinfo{author}{\bibfnamefont{X.-H.} \bibnamefont{Xia}},
	\bibinfo{journal}{J. Mat. Chem.} \textbf{\bibinfo{volume}{22}},
	\bibinfo{pages}{390} (\bibinfo{year}{2012}).
	
	\bibitem[{\citenamefont{Cocco et~al.}(2010)\citenamefont{Cocco, Cadelano, and
			Colombo}}]{cocco2010gap}
	\bibinfo{author}{\bibfnamefont{G.}~\bibnamefont{Cocco}},
	\bibinfo{author}{\bibfnamefont{E.}~\bibnamefont{Cadelano}}, \bibnamefont{and}
	\bibinfo{author}{\bibfnamefont{L.}~\bibnamefont{Colombo}},
	\bibinfo{journal}{Phys. Rev. B} \textbf{\bibinfo{volume}{81}},
	\bibinfo{pages}{241412} (\bibinfo{year}{2010}).
	
	\bibitem[{\citenamefont{Gui et~al.}(2008)\citenamefont{Gui, Li, and
			Zhong}}]{gui2008band}
	\bibinfo{author}{\bibfnamefont{G.}~\bibnamefont{Gui}},
	\bibinfo{author}{\bibfnamefont{J.}~\bibnamefont{Li}}, \bibnamefont{and}
	\bibinfo{author}{\bibfnamefont{J.}~\bibnamefont{Zhong}},
	\bibinfo{journal}{Phys. Rev. B} \textbf{\bibinfo{volume}{78}},
	\bibinfo{pages}{075435} (\bibinfo{year}{2008}).
	
	\bibitem[{\citenamefont{Rybkovskiy et~al.}(2014)\citenamefont{Rybkovskiy,
			Osadchy, and Obraztsova}}]{rybkovskiy2014transition}
	\bibinfo{author}{\bibfnamefont{D.~V.} \bibnamefont{Rybkovskiy}},
	\bibinfo{author}{\bibfnamefont{A.~V.} \bibnamefont{Osadchy}},
	\bibnamefont{and} \bibinfo{author}{\bibfnamefont{E.~D.}
		\bibnamefont{Obraztsova}}, \bibinfo{journal}{Phys. Rev. B}
	\textbf{\bibinfo{volume}{90}}, \bibinfo{pages}{235302}
	(\bibinfo{year}{2014}).
	
	\bibitem[{\citenamefont{Wickramaratne et~al.}(2014)\citenamefont{Wickramaratne,
			Zahid, and Lake}}]{wickramaratne2014electronic}
	\bibinfo{author}{\bibfnamefont{D.}~\bibnamefont{Wickramaratne}},
	\bibinfo{author}{\bibfnamefont{F.}~\bibnamefont{Zahid}}, \bibnamefont{and}
	\bibinfo{author}{\bibfnamefont{R.~K.} \bibnamefont{Lake}},
	\bibinfo{journal}{J. Chem. Phys} \textbf{\bibinfo{volume}{140}},
	\bibinfo{pages}{124710} (\bibinfo{year}{2014}).
	
	\bibitem[{\citenamefont{Cao et~al.}(2015)\citenamefont{Cao, Li, and
			Louie}}]{cao2015tunable}
	\bibinfo{author}{\bibfnamefont{T.}~\bibnamefont{Cao}},
	\bibinfo{author}{\bibfnamefont{Z.}~\bibnamefont{Li}}, \bibnamefont{and}
	\bibinfo{author}{\bibfnamefont{S.~G.} \bibnamefont{Louie}},
	\bibinfo{journal}{Phys. Rev. Lett.} \textbf{\bibinfo{volume}{114}},
	\bibinfo{pages}{236602} (\bibinfo{year}{2015}).
	
	\bibitem[{\citenamefont{Zólyomi et~al.}(2013)\citenamefont{Zólyomi, Drummond,
			and Fal'ko}}]{zolyomi_band_2013}
	\bibinfo{author}{\bibfnamefont{V.}~\bibnamefont{Zólyomi}},
	\bibinfo{author}{\bibfnamefont{N.~D.} \bibnamefont{Drummond}},
	\bibnamefont{and} \bibinfo{author}{\bibfnamefont{V.~I.}
		\bibnamefont{Fal'ko}}, \bibinfo{journal}{Phys. Rev. B}
	\textbf{\bibinfo{volume}{87}} (\bibinfo{year}{2013}).
	
	\bibitem[{\citenamefont{Wu et~al.}(2014)\citenamefont{Wu, Dai, Yu, Fan, Hu, and
			Yao}}]{wu2014magnetisms}
	\bibinfo{author}{\bibfnamefont{S.}~\bibnamefont{Wu}},
	\bibinfo{author}{\bibfnamefont{X.}~\bibnamefont{Dai}},
	\bibinfo{author}{\bibfnamefont{H.}~\bibnamefont{Yu}},
	\bibinfo{author}{\bibfnamefont{H.}~\bibnamefont{Fan}},
	\bibinfo{author}{\bibfnamefont{J.}~\bibnamefont{Hu}}, \bibnamefont{and}
	\bibinfo{author}{\bibfnamefont{W.}~\bibnamefont{Yao}},
	\bibinfo{journal}{arXiv preprint arXiv:1409.4733}  (\bibinfo{year}{2014}).
	
	\bibitem[{\citenamefont{Magorrian et~al.}(2016)\citenamefont{Magorrian,
			Z{\'o}lyomi, and Fal'ko}}]{magorrian2016electronic}
	\bibinfo{author}{\bibfnamefont{S.}~\bibnamefont{Magorrian}},
	\bibinfo{author}{\bibfnamefont{V.}~\bibnamefont{Z{\'o}lyomi}},
	\bibnamefont{and} \bibinfo{author}{\bibfnamefont{V.}~\bibnamefont{Fal'ko}},
	\bibinfo{journal}{Phys. Rev. B} \textbf{\bibinfo{volume}{94}},
	\bibinfo{pages}{245431} (\bibinfo{year}{2016}).
	
	\bibitem[{\citenamefont{Lifshitz et~al.}(1960)}]{lifshitz1960anomalies}
	\bibinfo{author}{\bibfnamefont{I.}~\bibnamefont{Lifshitz}}
	\bibnamefont{et~al.}, \bibinfo{journal}{Sov. Phys. JETP}
	\textbf{\bibinfo{volume}{11}}, \bibinfo{pages}{1130} (\bibinfo{year}{1960}).
	
	\bibitem[{\citenamefont{Wickramaratne et~al.}(2015)\citenamefont{Wickramaratne,
			Zahid, and Lake}}]{wickramaratne2015electronic}
	\bibinfo{author}{\bibfnamefont{D.}~\bibnamefont{Wickramaratne}},
	\bibinfo{author}{\bibfnamefont{F.}~\bibnamefont{Zahid}}, \bibnamefont{and}
	\bibinfo{author}{\bibfnamefont{R.~K.} \bibnamefont{Lake}},
	\bibinfo{journal}{J. Appl. Phys.} \textbf{\bibinfo{volume}{118}},
	\bibinfo{pages}{075101} (\bibinfo{year}{2015}).
	
	\bibitem[{\citenamefont{Do et~al.}(2015)\citenamefont{Do, Mahanti, and
			Lai}}]{do2015spin}
	\bibinfo{author}{\bibfnamefont{D.~T.} \bibnamefont{Do}},
	\bibinfo{author}{\bibfnamefont{S.~D.} \bibnamefont{Mahanti}},
	\bibnamefont{and} \bibinfo{author}{\bibfnamefont{C.~W.} \bibnamefont{Lai}},
	\bibinfo{journal}{Sci. Rep.} \textbf{\bibinfo{volume}{5}},
	\bibinfo{pages}{17044} (\bibinfo{year}{2015}).
	
	\bibitem[{\citenamefont{Bychkov and Rashba}(1984)}]{bychkov1984properties}
	\bibinfo{author}{\bibfnamefont{Y.~A.} \bibnamefont{Bychkov}} \bibnamefont{and}
	\bibinfo{author}{\bibfnamefont{E.~I.} \bibnamefont{Rashba}},
	\bibinfo{journal}{JETP lett} \textbf{\bibinfo{volume}{39}},
	\bibinfo{pages}{78} (\bibinfo{year}{1984}).
	
	\bibitem[{\citenamefont{Datta and Das}(1990)}]{datta1990electronic}
	\bibinfo{author}{\bibfnamefont{S.}~\bibnamefont{Datta}} \bibnamefont{and}
	\bibinfo{author}{\bibfnamefont{B.}~\bibnamefont{Das}},
	\bibinfo{journal}{Appl. Phys. Lett.} \textbf{\bibinfo{volume}{56}},
	\bibinfo{pages}{665} (\bibinfo{year}{1990}).
	
	\bibitem[{\citenamefont{Absor et~al.}(2018{\natexlab{a}})\citenamefont{Absor,
			Kotaka, Ishii, and Saito}}]{absor2018tunable}
	\bibinfo{author}{\bibfnamefont{M.~A.~U.} \bibnamefont{Absor}},
	\bibinfo{author}{\bibfnamefont{H.}~\bibnamefont{Kotaka}},
	\bibinfo{author}{\bibfnamefont{F.}~\bibnamefont{Ishii}}, \bibnamefont{and}
	\bibinfo{author}{\bibfnamefont{M.}~\bibnamefont{Saito}},
	\bibinfo{journal}{Jpn. J. Appl. Phys.} \textbf{\bibinfo{volume}{57}},
	\bibinfo{pages}{04FP01} (\bibinfo{year}{2018}{\natexlab{a}}).
	
	\bibitem[{\citenamefont{Zhang and
			Schwingenschl{\"o}gl}(2018)}]{zhang2018rashba}
	\bibinfo{author}{\bibfnamefont{Q.}~\bibnamefont{Zhang}} \bibnamefont{and}
	\bibinfo{author}{\bibfnamefont{U.}~\bibnamefont{Schwingenschl{\"o}gl}},
	\bibinfo{journal}{Phys. Rev. B} \textbf{\bibinfo{volume}{97}},
	\bibinfo{pages}{155415} (\bibinfo{year}{2018}).
	
	\bibitem[{\citenamefont{Premasiri et~al.}(2018)\citenamefont{Premasiri, Kumar,
			Sucharitakul, Ulaganathan, Sankar, Chou, Chen, and
			Gao}}]{premasiri2018tuning}
	\bibinfo{author}{\bibfnamefont{K.}~\bibnamefont{Premasiri}},
	\bibinfo{author}{\bibfnamefont{S.}~\bibnamefont{Kumar}},
	\bibinfo{author}{\bibfnamefont{S.}~\bibnamefont{Sucharitakul}},
	\bibinfo{author}{\bibfnamefont{R.~K.} \bibnamefont{Ulaganathan}},
	\bibinfo{author}{\bibfnamefont{R.}~\bibnamefont{Sankar}},
	\bibinfo{author}{\bibfnamefont{F.-C.} \bibnamefont{Chou}},
	\bibinfo{author}{\bibfnamefont{Y.-T.} \bibnamefont{Chen}}, \bibnamefont{and}
	\bibinfo{author}{\bibfnamefont{X.~P.} \bibnamefont{Gao}},
	\bibinfo{journal}{Nano Lett.}  (\bibinfo{year}{2018}).
	
	\bibitem[{\citenamefont{Nitta et~al.}(1997)\citenamefont{Nitta, Akazaki,
			Takayanagi, and Enoki}}]{nitta1997gate}
	\bibinfo{author}{\bibfnamefont{J.}~\bibnamefont{Nitta}},
	\bibinfo{author}{\bibfnamefont{T.}~\bibnamefont{Akazaki}},
	\bibinfo{author}{\bibfnamefont{H.}~\bibnamefont{Takayanagi}},
	\bibnamefont{and} \bibinfo{author}{\bibfnamefont{T.}~\bibnamefont{Enoki}},
	\bibinfo{journal}{Phys. Rev. Lett.} \textbf{\bibinfo{volume}{78}},
	\bibinfo{pages}{1335} (\bibinfo{year}{1997}).
	
	\bibitem[{\citenamefont{LaShell}(1996)}]{lashell1996s}
	\bibinfo{author}{\bibfnamefont{S.}~\bibnamefont{LaShell}},
	\bibinfo{journal}{Phys. Rev. Lett.} \textbf{\bibinfo{volume}{77}},
	\bibinfo{pages}{3419} (\bibinfo{year}{1996}).
	
	\bibitem[{\citenamefont{{\v{Z}}uti{\'c}}(2004)}]{vzutic2004vzutic}
	\bibinfo{author}{\bibfnamefont{I.}~\bibnamefont{{\v{Z}}uti{\'c}}},
	\bibinfo{journal}{Rev. Mod. Phys.} \textbf{\bibinfo{volume}{76}},
	\bibinfo{pages}{323} (\bibinfo{year}{2004}).
	
	\bibitem[{\citenamefont{Kimura et~al.}(2010)\citenamefont{Kimura, Krasovskii,
			Nishimura, Miyamoto, Kadono, Kanomaru, Chulkov, Bihlmayer, Shimada, Namatame
			et~al.}}]{kimura2010strong}
	\bibinfo{author}{\bibfnamefont{A.}~\bibnamefont{Kimura}},
	\bibinfo{author}{\bibfnamefont{E.}~\bibnamefont{Krasovskii}},
	\bibinfo{author}{\bibfnamefont{R.}~\bibnamefont{Nishimura}},
	\bibinfo{author}{\bibfnamefont{K.}~\bibnamefont{Miyamoto}},
	\bibinfo{author}{\bibfnamefont{T.}~\bibnamefont{Kadono}},
	\bibinfo{author}{\bibfnamefont{K.}~\bibnamefont{Kanomaru}},
	\bibinfo{author}{\bibfnamefont{E.}~\bibnamefont{Chulkov}},
	\bibinfo{author}{\bibfnamefont{G.}~\bibnamefont{Bihlmayer}},
	\bibinfo{author}{\bibfnamefont{K.}~\bibnamefont{Shimada}},
	\bibinfo{author}{\bibfnamefont{H.}~\bibnamefont{Namatame}},
	\bibnamefont{et~al.}, \bibinfo{journal}{Phys. Rev. Lett.}
	\textbf{\bibinfo{volume}{105}}, \bibinfo{pages}{076804}
	(\bibinfo{year}{2010}).
	
	\bibitem[{\citenamefont{Konschuh et~al.}(2010)\citenamefont{Konschuh, Gmitra,
			and Fabian}}]{konschuh2010tight}
	\bibinfo{author}{\bibfnamefont{S.}~\bibnamefont{Konschuh}},
	\bibinfo{author}{\bibfnamefont{M.}~\bibnamefont{Gmitra}}, \bibnamefont{and}
	\bibinfo{author}{\bibfnamefont{J.}~\bibnamefont{Fabian}},
	\bibinfo{journal}{Phys. Rev. B} \textbf{\bibinfo{volume}{82}},
	\bibinfo{pages}{245412} (\bibinfo{year}{2010}).
	
	\bibitem[{\citenamefont{Cheng et~al.}(2013)\citenamefont{Cheng, Zhu, Tahir, and
			Schwingenschl{\"o}gl}}]{cheng2013spin}
	\bibinfo{author}{\bibfnamefont{Y.}~\bibnamefont{Cheng}},
	\bibinfo{author}{\bibfnamefont{Z.}~\bibnamefont{Zhu}},
	\bibinfo{author}{\bibfnamefont{M.}~\bibnamefont{Tahir}}, \bibnamefont{and}
	\bibinfo{author}{\bibfnamefont{U.}~\bibnamefont{Schwingenschl{\"o}gl}},
	\bibinfo{journal}{Europhys. Lett.} \textbf{\bibinfo{volume}{102}},
	\bibinfo{pages}{57001} (\bibinfo{year}{2013}).
	
	\bibitem[{\citenamefont{Liu et~al.}(2013)\citenamefont{Liu, Guo, and
			Freeman}}]{liu2013tunable}
	\bibinfo{author}{\bibfnamefont{Q.}~\bibnamefont{Liu}},
	\bibinfo{author}{\bibfnamefont{Y.}~\bibnamefont{Guo}}, \bibnamefont{and}
	\bibinfo{author}{\bibfnamefont{A.~J.} \bibnamefont{Freeman}},
	\bibinfo{journal}{Nano Lett.} \textbf{\bibinfo{volume}{13}},
	\bibinfo{pages}{5264} (\bibinfo{year}{2013}).
	
	\bibitem[{\citenamefont{Min et~al.}(2006)\citenamefont{Min, Hill, Sinitsyn,
			Sahu, Kleinman, and MacDonald}}]{min2006intrinsic}
	\bibinfo{author}{\bibfnamefont{H.}~\bibnamefont{Min}},
	\bibinfo{author}{\bibfnamefont{J.}~\bibnamefont{Hill}},
	\bibinfo{author}{\bibfnamefont{N.~A.} \bibnamefont{Sinitsyn}},
	\bibinfo{author}{\bibfnamefont{B.}~\bibnamefont{Sahu}},
	\bibinfo{author}{\bibfnamefont{L.}~\bibnamefont{Kleinman}}, \bibnamefont{and}
	\bibinfo{author}{\bibfnamefont{A.~H.} \bibnamefont{MacDonald}},
	\bibinfo{journal}{Phys. Rev. B} \textbf{\bibinfo{volume}{74}},
	\bibinfo{pages}{165310} (\bibinfo{year}{2006}).
	
	\bibitem[{\citenamefont{Ishizaka}(2011)}]{ishizaka2011k}
	\bibinfo{author}{\bibfnamefont{K.}~\bibnamefont{Ishizaka}},
	\bibinfo{journal}{Nat. Mater.} \textbf{\bibinfo{volume}{10}},
	\bibinfo{pages}{521} (\bibinfo{year}{2011}).
	
	\bibitem[{\citenamefont{Ast et~al.}(2007)\citenamefont{Ast, Henk, Ernst,
			Moreschini, Falub, Pacil{\'e}, Bruno, Kern, and Grioni}}]{ast2007giant}
	\bibinfo{author}{\bibfnamefont{C.~R.} \bibnamefont{Ast}},
	\bibinfo{author}{\bibfnamefont{J.}~\bibnamefont{Henk}},
	\bibinfo{author}{\bibfnamefont{A.}~\bibnamefont{Ernst}},
	\bibinfo{author}{\bibfnamefont{L.}~\bibnamefont{Moreschini}},
	\bibinfo{author}{\bibfnamefont{M.~C.} \bibnamefont{Falub}},
	\bibinfo{author}{\bibfnamefont{D.}~\bibnamefont{Pacil{\'e}}},
	\bibinfo{author}{\bibfnamefont{P.}~\bibnamefont{Bruno}},
	\bibinfo{author}{\bibfnamefont{K.}~\bibnamefont{Kern}}, \bibnamefont{and}
	\bibinfo{author}{\bibfnamefont{M.}~\bibnamefont{Grioni}},
	\bibinfo{journal}{Phys. Rev. Lett.} \textbf{\bibinfo{volume}{98}},
	\bibinfo{pages}{186807} (\bibinfo{year}{2007}).
	
	\bibitem[{\citenamefont{Koroteev et~al.}(2004)\citenamefont{Koroteev,
			Bihlmayer, Gayone, Chulkov, Bl{\"u}gel, Echenique, and
			Hofmann}}]{koroteev2004strong}
	\bibinfo{author}{\bibfnamefont{Y.~M.} \bibnamefont{Koroteev}},
	\bibinfo{author}{\bibfnamefont{G.}~\bibnamefont{Bihlmayer}},
	\bibinfo{author}{\bibfnamefont{J.}~\bibnamefont{Gayone}},
	\bibinfo{author}{\bibfnamefont{E.}~\bibnamefont{Chulkov}},
	\bibinfo{author}{\bibfnamefont{S.}~\bibnamefont{Bl{\"u}gel}},
	\bibinfo{author}{\bibfnamefont{P.~M.} \bibnamefont{Echenique}},
	\bibnamefont{and} \bibinfo{author}{\bibfnamefont{P.}~\bibnamefont{Hofmann}},
	\bibinfo{journal}{Phys. Rev. Lett.} \textbf{\bibinfo{volume}{93}},
	\bibinfo{pages}{046403} (\bibinfo{year}{2004}).
	
	\bibitem[{\citenamefont{Lee et~al.}(2015)\citenamefont{Lee, Yun, and
			Lee}}]{lee2015giant}
	\bibinfo{author}{\bibfnamefont{K.}~\bibnamefont{Lee}},
	\bibinfo{author}{\bibfnamefont{W.~S.} \bibnamefont{Yun}}, \bibnamefont{and}
	\bibinfo{author}{\bibfnamefont{J.}~\bibnamefont{Lee}},
	\bibinfo{journal}{Phys. Rev. B} \textbf{\bibinfo{volume}{91}},
	\bibinfo{pages}{125420} (\bibinfo{year}{2015}).
	
	\bibitem[{\citenamefont{{\v{Z}}uti{\'c}
			et~al.}(2004)\citenamefont{{\v{Z}}uti{\'c}, Fabian, and
			Sarma}}]{gui2008band11}
	\bibinfo{author}{\bibfnamefont{I.}~\bibnamefont{{\v{Z}}uti{\'c}}},
	\bibinfo{author}{\bibfnamefont{J.}~\bibnamefont{Fabian}}, \bibnamefont{and}
	\bibinfo{author}{\bibfnamefont{S.~D.} \bibnamefont{Sarma}},
	\bibinfo{journal}{Rev. Mod. Phys.} \textbf{\bibinfo{volume}{76}},
	\bibinfo{pages}{323} (\bibinfo{year}{2004}).
	
	\bibitem[{\citenamefont{Zhang et~al.}(2015)\citenamefont{Zhang, Yan, Li, Chen,
			and Zeng}}]{zhang2015atomically}
	\bibinfo{author}{\bibfnamefont{S.}~\bibnamefont{Zhang}},
	\bibinfo{author}{\bibfnamefont{Z.}~\bibnamefont{Yan}},
	\bibinfo{author}{\bibfnamefont{Y.}~\bibnamefont{Li}},
	\bibinfo{author}{\bibfnamefont{Z.}~\bibnamefont{Chen}}, \bibnamefont{and}
	\bibinfo{author}{\bibfnamefont{H.}~\bibnamefont{Zeng}},
	\bibinfo{journal}{Angewandte Chemie} \textbf{\bibinfo{volume}{127}},
	\bibinfo{pages}{3155} (\bibinfo{year}{2015}).
	
	\bibitem[{\citenamefont{Giannozzi et~al.}(2009)\citenamefont{Giannozzi, Baroni,
			Bonini, Calandra, Car, Cavazzoni, Ceresoli, Chiarotti, Cococcioni, Dabo
			et~al.}}]{giannozzi2009quantum}
	\bibinfo{author}{\bibfnamefont{P.}~\bibnamefont{Giannozzi}},
	\bibinfo{author}{\bibfnamefont{S.}~\bibnamefont{Baroni}},
	\bibinfo{author}{\bibfnamefont{N.}~\bibnamefont{Bonini}},
	\bibinfo{author}{\bibfnamefont{M.}~\bibnamefont{Calandra}},
	\bibinfo{author}{\bibfnamefont{R.}~\bibnamefont{Car}},
	\bibinfo{author}{\bibfnamefont{C.}~\bibnamefont{Cavazzoni}},
	\bibinfo{author}{\bibfnamefont{D.}~\bibnamefont{Ceresoli}},
	\bibinfo{author}{\bibfnamefont{G.~L.} \bibnamefont{Chiarotti}},
	\bibinfo{author}{\bibfnamefont{M.}~\bibnamefont{Cococcioni}},
	\bibinfo{author}{\bibfnamefont{I.}~\bibnamefont{Dabo}}, \bibnamefont{et~al.},
	\bibinfo{journal}{J. Phys.:Condensed Matter} \textbf{\bibinfo{volume}{21}},
	\bibinfo{pages}{395502} (\bibinfo{year}{2009}).
	
	\bibitem[{\citenamefont{Giannozzi et~al.}(2017)\citenamefont{Giannozzi,
			Andreussi, Brumme, Bunau, Nardelli, Calandra, Car, Cavazzoni, Ceresoli,
			Cococcioni et~al.}}]{giannozzi2017advanced}
	\bibinfo{author}{\bibfnamefont{P.}~\bibnamefont{Giannozzi}},
	\bibinfo{author}{\bibfnamefont{O.}~\bibnamefont{Andreussi}},
	\bibinfo{author}{\bibfnamefont{T.}~\bibnamefont{Brumme}},
	\bibinfo{author}{\bibfnamefont{O.}~\bibnamefont{Bunau}},
	\bibinfo{author}{\bibfnamefont{M.~B.} \bibnamefont{Nardelli}},
	\bibinfo{author}{\bibfnamefont{M.}~\bibnamefont{Calandra}},
	\bibinfo{author}{\bibfnamefont{R.}~\bibnamefont{Car}},
	\bibinfo{author}{\bibfnamefont{C.}~\bibnamefont{Cavazzoni}},
	\bibinfo{author}{\bibfnamefont{D.}~\bibnamefont{Ceresoli}},
	\bibinfo{author}{\bibfnamefont{M.}~\bibnamefont{Cococcioni}},
	\bibnamefont{et~al.}, \bibinfo{journal}{J. Phys.:Condensed Matter}
	\textbf{\bibinfo{volume}{29}}, \bibinfo{pages}{465901}
	(\bibinfo{year}{2017}).
	
	\bibitem[{\citenamefont{Bl{\"o}chl}(1994)}]{blochl94}
	\bibinfo{author}{\bibfnamefont{P.~E.} \bibnamefont{Bl{\"o}chl}},
	\bibinfo{journal}{Phys. Rev. B} \textbf{\bibinfo{volume}{50}},
	\bibinfo{pages}{17953} (\bibinfo{year}{1994}).
	
	\bibitem[{\citenamefont{Kresse and Joubert}(1999)}]{kresse99}
	\bibinfo{author}{\bibfnamefont{G.}~\bibnamefont{Kresse}} \bibnamefont{and}
	\bibinfo{author}{\bibfnamefont{D.}~\bibnamefont{Joubert}},
	\bibinfo{journal}{Phys. Rev. B} \textbf{\bibinfo{volume}{59}},
	\bibinfo{pages}{1758} (\bibinfo{year}{1999}).
	
	\bibitem[{\citenamefont{Perdew et~al.}(2008)\citenamefont{Perdew, Ruzsinszky,
			Csonka, Vydrov, Scuseria, Constantin, Zhou, and Burke}}]{perdew2008restoring}
	\bibinfo{author}{\bibfnamefont{J.~P.} \bibnamefont{Perdew}},
	\bibinfo{author}{\bibfnamefont{A.}~\bibnamefont{Ruzsinszky}},
	\bibinfo{author}{\bibfnamefont{G.~I.} \bibnamefont{Csonka}},
	\bibinfo{author}{\bibfnamefont{O.~A.} \bibnamefont{Vydrov}},
	\bibinfo{author}{\bibfnamefont{G.~E.} \bibnamefont{Scuseria}},
	\bibinfo{author}{\bibfnamefont{L.~A.} \bibnamefont{Constantin}},
	\bibinfo{author}{\bibfnamefont{X.}~\bibnamefont{Zhou}}, \bibnamefont{and}
	\bibinfo{author}{\bibfnamefont{K.}~\bibnamefont{Burke}},
	\bibinfo{journal}{Phys. Rev. Lett.} \textbf{\bibinfo{volume}{100}},
	\bibinfo{pages}{136406} (\bibinfo{year}{2008}).
	
	\bibitem[{\citenamefont{Ge et~al.}(2017)\citenamefont{Ge, Qin, Yao, and
			L{\"u}}}]{ge2017first}
	\bibinfo{author}{\bibfnamefont{X.-J.} \bibnamefont{Ge}},
	\bibinfo{author}{\bibfnamefont{D.}~\bibnamefont{Qin}},
	\bibinfo{author}{\bibfnamefont{K.-L.} \bibnamefont{Yao}}, \bibnamefont{and}
	\bibinfo{author}{\bibfnamefont{J.-T.} \bibnamefont{L{\"u}}},
	\bibinfo{journal}{J. Phys. D: Appl. Phys.} \textbf{\bibinfo{volume}{50}},
	\bibinfo{pages}{405301} (\bibinfo{year}{2017}).
	
	\bibitem[{\citenamefont{Cai et~al.}(2016{\natexlab{a}})\citenamefont{Cai, Kang,
			Sahin, Chen, Suslu, Wu, Peeters, Meng, and Tongay}}]{cai2016exciton}
	\bibinfo{author}{\bibfnamefont{H.}~\bibnamefont{Cai}},
	\bibinfo{author}{\bibfnamefont{J.}~\bibnamefont{Kang}},
	\bibinfo{author}{\bibfnamefont{H.}~\bibnamefont{Sahin}},
	\bibinfo{author}{\bibfnamefont{B.}~\bibnamefont{Chen}},
	\bibinfo{author}{\bibfnamefont{A.}~\bibnamefont{Suslu}},
	\bibinfo{author}{\bibfnamefont{K.}~\bibnamefont{Wu}},
	\bibinfo{author}{\bibfnamefont{F.}~\bibnamefont{Peeters}},
	\bibinfo{author}{\bibfnamefont{X.}~\bibnamefont{Meng}}, \bibnamefont{and}
	\bibinfo{author}{\bibfnamefont{S.}~\bibnamefont{Tongay}},
	\bibinfo{journal}{Nanotechnology} \textbf{\bibinfo{volume}{27}},
	\bibinfo{pages}{065203} (\bibinfo{year}{2016}{\natexlab{a}}).
	
	\bibitem[{\citenamefont{Cai et~al.}(2016{\natexlab{b}})\citenamefont{Cai,
			Soignard, Ataca, Chen, Ko, Aoki, Pant, Meng, Yang, Grossman
			et~al.}}]{cai2016band}
	\bibinfo{author}{\bibfnamefont{H.}~\bibnamefont{Cai}},
	\bibinfo{author}{\bibfnamefont{E.}~\bibnamefont{Soignard}},
	\bibinfo{author}{\bibfnamefont{C.}~\bibnamefont{Ataca}},
	\bibinfo{author}{\bibfnamefont{B.}~\bibnamefont{Chen}},
	\bibinfo{author}{\bibfnamefont{C.}~\bibnamefont{Ko}},
	\bibinfo{author}{\bibfnamefont{T.}~\bibnamefont{Aoki}},
	\bibinfo{author}{\bibfnamefont{A.}~\bibnamefont{Pant}},
	\bibinfo{author}{\bibfnamefont{X.}~\bibnamefont{Meng}},
	\bibinfo{author}{\bibfnamefont{S.}~\bibnamefont{Yang}},
	\bibinfo{author}{\bibfnamefont{J.}~\bibnamefont{Grossman}},
	\bibnamefont{et~al.}, \bibinfo{journal}{Adv. Mater.}
	\textbf{\bibinfo{volume}{28}}, \bibinfo{pages}{7375}
	(\bibinfo{year}{2016}{\natexlab{b}}).
	
	\bibitem[{\citenamefont{Huang et~al.}(2015)\citenamefont{Huang, Chen, and
			Li}}]{huang2015effects}
	\bibinfo{author}{\bibfnamefont{L.}~\bibnamefont{Huang}},
	\bibinfo{author}{\bibfnamefont{Z.}~\bibnamefont{Chen}}, \bibnamefont{and}
	\bibinfo{author}{\bibfnamefont{J.}~\bibnamefont{Li}}, \bibinfo{journal}{RSC
		Advances} \textbf{\bibinfo{volume}{5}}, \bibinfo{pages}{5788}
	(\bibinfo{year}{2015}).
	
	\bibitem[{\citenamefont{Zhuang and Hennig}(2013)}]{zhuang2013single}
	\bibinfo{author}{\bibfnamefont{H.~L.} \bibnamefont{Zhuang}} \bibnamefont{and}
	\bibinfo{author}{\bibfnamefont{R.~G.} \bibnamefont{Hennig}},
	\bibinfo{journal}{Chem. Mater.} \textbf{\bibinfo{volume}{25}},
	\bibinfo{pages}{3232} (\bibinfo{year}{2013}).
	
	\bibitem[{\citenamefont{Absor et~al.}(2018{\natexlab{b}})\citenamefont{Absor,
			Santoso, Abraha, Kotaka, Ishii, Saito et~al.}}]{absor2018strong}
	\bibinfo{author}{\bibfnamefont{M.~A.~U.} \bibnamefont{Absor}},
	\bibinfo{author}{\bibfnamefont{I.}~\bibnamefont{Santoso}},
	\bibinfo{author}{\bibfnamefont{K.}~\bibnamefont{Abraha}},
	\bibinfo{author}{\bibfnamefont{H.}~\bibnamefont{Kotaka}},
	\bibinfo{author}{\bibfnamefont{F.}~\bibnamefont{Ishii}},
	\bibinfo{author}{\bibfnamefont{M.}~\bibnamefont{Saito}},
	\bibnamefont{et~al.}, \bibinfo{journal}{Phys. Rev. B}
	\textbf{\bibinfo{volume}{97}}, \bibinfo{pages}{205138}
	(\bibinfo{year}{2018}{\natexlab{b}}).
	
	\bibitem[{\citenamefont{Ma et~al.}(2014)\citenamefont{Ma, Dai, Wei, Li, and
			Huang}}]{ma2014emergence}
	\bibinfo{author}{\bibfnamefont{Y.}~\bibnamefont{Ma}},
	\bibinfo{author}{\bibfnamefont{Y.}~\bibnamefont{Dai}},
	\bibinfo{author}{\bibfnamefont{W.}~\bibnamefont{Wei}},
	\bibinfo{author}{\bibfnamefont{X.}~\bibnamefont{Li}}, \bibnamefont{and}
	\bibinfo{author}{\bibfnamefont{B.}~\bibnamefont{Huang}},
	\bibinfo{journal}{Phys. Chem. Chem. Phys.} \textbf{\bibinfo{volume}{16}},
	\bibinfo{pages}{17603} (\bibinfo{year}{2014}).
	
	\bibitem[{\citenamefont{Li et~al.}(2017)\citenamefont{Li, Wei, Zhao, Huang, and
			Dai}}]{li2017electronic}
	\bibinfo{author}{\bibfnamefont{F.}~\bibnamefont{Li}},
	\bibinfo{author}{\bibfnamefont{W.}~\bibnamefont{Wei}},
	\bibinfo{author}{\bibfnamefont{P.}~\bibnamefont{Zhao}},
	\bibinfo{author}{\bibfnamefont{B.}~\bibnamefont{Huang}}, \bibnamefont{and}
	\bibinfo{author}{\bibfnamefont{Y.}~\bibnamefont{Dai}}, \bibinfo{journal}{J.
		Phys. Chem. Lett} \textbf{\bibinfo{volume}{8}}, \bibinfo{pages}{5959}
	(\bibinfo{year}{2017}).
	
	\bibitem[{\citenamefont{Yao et~al.}(2017)\citenamefont{Yao, Cai, Tong, Gong,
			Wang, Wan, Duan, and Chu}}]{yao2017manipulation}
	\bibinfo{author}{\bibfnamefont{Q.-F.} \bibnamefont{Yao}},
	\bibinfo{author}{\bibfnamefont{J.}~\bibnamefont{Cai}},
	\bibinfo{author}{\bibfnamefont{W.-Y.} \bibnamefont{Tong}},
	\bibinfo{author}{\bibfnamefont{S.-J.} \bibnamefont{Gong}},
	\bibinfo{author}{\bibfnamefont{J.-Q.} \bibnamefont{Wang}},
	\bibinfo{author}{\bibfnamefont{X.}~\bibnamefont{Wan}},
	\bibinfo{author}{\bibfnamefont{C.-G.} \bibnamefont{Duan}}, \bibnamefont{and}
	\bibinfo{author}{\bibfnamefont{J.}~\bibnamefont{Chu}},
	\bibinfo{journal}{Phys. Rev. B} \textbf{\bibinfo{volume}{95}},
	\bibinfo{pages}{165401} (\bibinfo{year}{2017}).
	
	\bibitem[{\citenamefont{Singh and Romero}(2017)}]{singh2017giant}
	\bibinfo{author}{\bibfnamefont{S.}~\bibnamefont{Singh}} \bibnamefont{and}
	\bibinfo{author}{\bibfnamefont{A.~H.} \bibnamefont{Romero}},
	\bibinfo{journal}{Phys. Rev. B} \textbf{\bibinfo{volume}{95}},
	\bibinfo{pages}{165444} (\bibinfo{year}{2017}).
	
\end{thebibliography}

\end{document}